\documentclass[twocolumn,preprintnumbers,jcp,showpacs,superscriptaddress]{revtex4-1}

\usepackage{color}
\usepackage{palatino}
\usepackage{dcolumn}
\usepackage[latin1]{inputenc}
\usepackage{epsf}
\usepackage{amsmath,amssymb,amscd,amsthm}
\usepackage{latexsym}
\usepackage{calc}
\usepackage[usenames,dvipsnames]{xcolor}
\usepackage{epsfig}
\usepackage[pdftex, pdfstartview = {FitH}]{hyperref}
\usepackage{graphicx}
\usepackage{amsfonts}
\usepackage{appendix}
\usepackage{lastpage}
\usepackage{fancyhdr}
\usepackage{enumerate}

\usepackage{bm}

\def\bA{{\bf A}}
\def\bP{{\bf P}}

\begin{document}
\title{Nuclear velocity perturbation theory for vibrational circular dichroism: An approach based on the exact factorization of the electron-nuclear wave function}

\author{Arne Scherrer,$^{a,c,d,*}$ Federica Agostini,$^{b,*}$ Daniel Sebastiani,$^a$ E. K. U. Gross$^b$ and Rodolphe Vuilleumier$^{c,d}$\\
\vspace{6pt} $^{a}${\em{Martin-Luther-University Halle-Wittenberg, von-Danckelmann-Platz 4, D-06120 Halle, Germany}}\\
\vspace{6pt} $^{b}${\em{Max-Planck-Institute of Microstructure Physics, Weinberg 2, D-06120 Halle, Germany}}\\
\vspace{6pt} $^{c}${\em{UMR 8640 ENS-CNRS-UPMC, D\'epartement de Chimie, 24 rue Lhomond, \'Ecole Normale Sup\'erieure, 75005 Paris, France}}\\
\vspace{6pt} $^{d}${\em{UPMC Universit\'e Paris 06, 4, Place Jussieu, 75005 Paris, France}}\\
\vspace{6pt} $^{*}$These authors contributed equally.}

\date{\today}
\pacs{}
\begin{abstract}
The nuclear velocity perturbation theory (NVPT) for vibrational circular dichroism (VCD) is derived from the exact factorization of the electron-nuclear wave function. This new formalism offers an \textit{exact} starting point to include correction terms to the Born-Oppenheimer (BO) form of the molecular wave function, similarly to the complete-adiabatic approximation. The corrections depend on a small parameter that, in a classical treatment of the nuclei, is identified as the nuclear velocity. Apart from proposing a rigorous basis for the NVPT, we show that the rotational strengths, related to the intensity of the VCD signal, contain a new contribution beyond-BO that can be evaluated with the NVPT and that only arises when the exact factorization approach is employed. Numerical results are presented for chiral and non-chiral systems to test the validity of the approach.
\end{abstract}
\maketitle 

\section{Introduction}\label{sec: intro}
Vibrational circular dichroism (VCD)~\cite{Nafie_ARPC1997, Vass_AR2011, Nafie} in molecules refers to the difference in absorption of left and right circularly polarized light in the infrared region of the electromagnetic spectrum. In contrast to circular dichroism, that originates in electronic transitions, VCD is the difference in interaction of a molecule with radiation of opposite circular polarizations when it undergoes vibrational transitions. Experimentally, VCD is employed to probe the absolute configuration of chiral molecules in solution and provides detailed structural information, thus being a very sensitive form of vibrational spectroscopy.

From the theoretical point of view~\cite{Barone_JCP2012, Barone_JPCL2012, ruud2012ab, Chabay_JCP1972, Stephens_C2000, Stephens_JPC1994, Schellman_JCP1973, *Schellman_JCP1974, Snir_B1975, Cross_1955, Faulkner_JACS1977, Thirunamachandran_MP1978, Keiderling_JACS1977, Walnut_CPL1977, Buckingham_CP1978, Freedman_JCP1983, Nafie_JCP1983, Stephens_JPC1985, Nafie_JCP1992}, the Born-Oppenheimer (BO)~\cite{BO_AP1927}, or adiabatic, treatment of the coupled motion of electrons and nuclei in molecular systems is inadequate for predicting VCD.
Since the intensity of the VCD signal is proportional to the rotational strength for a transition between two vibrational states, the calculation of the electric current and of the magnetic dipole moment (and of their scalar product) is required. The electric current and the magnetic dipole moment contain both electronic and nuclear contributions, but when the BO approximation is employed, the electronic contributions identically vanish. This is due to the fact that the ground-state electronic wave function is real for a non-degenerate adiabatic state and therefore the expectation values of the purely imaginary (Hermitian) electric current~\cite{Nafie_JPCA1997, Nafie_JACS1997, Nafie_JPCA1998, Barth_CPL2009, Schild_CP2010, Paulus_JPCB2011} and magnetic dipole moment operators vanish~\cite{Freedman_JCP1983}.
Therefore, VCD appears a fundamentally non-adiabatic (beyond-BO) process, thus requiring a theoretical approach able to explicitly treat the dynamical coupling between electronic and nuclear degrees freedom in molecules.

A practical question~\cite{Patchkovskii_JCP2012} arises at this point, as to whether such coupling can be accounted for within a standard ab-initio molecular dynamics formulation. Among the most successful ideas are in fact those resorting to the treatment of beyond-BO effects as a perturbation to the BO problem, numerically less expensive than a full non-adiabatic calculation but indeed not consistent if strong non-adiabatic effects are expected, e.g. in the presence of conical intersections. Examples are the approaches proposed by Nafie~\cite{Nafie_JCP1983}, employing the complete-adiabatic expression of the electron-nuclear wave function, and by Stephens~\cite{Stephens_JPC1985}, introducing the magnetic field perturbation theory. These methods allow to overcome the problems encountered in the BO calculation of VCD, while exploiting the advantages of the BO formalism like the product form of the electron-nuclear wave function. Recently, VCD has been calculated by developing and implementing a nuclear velocity perturbation theory (NVPT)~\cite{Scherrer_JCTC2013} based on the complete-adiabatic approach of Nafie~\cite{Nafie_JCP1983}. In this formulation, non-adiabatic corrections to the electronic adiabatic ground-state are perturbatively taken into account and are induced by a ``small'' nuclear velocity.

In this paper we propose a novel approach to NVPT, based on the exact factorization of the electron-nuclear wave function~\cite{Gross_PRL2010, Gross_JCP2012}. The advantage of this formulation comes from using a product form, like in the BO approximation, of the wave function, that is not the result of an approximation but an \textit{exact} starting point. The electron-nuclear wave function is written as a single product of an electronic many-body factor, parametrically depending on the nuclear positions, and a nuclear wave function. The latter can be interpreted as a proper nuclear wave function since it leads to the exact nuclear density and current-density. Moreover, when the product form is inserted into the time-dependent Schr\"odinger equation (TDSE), two coupled equations for the components of the full wave function are derived, with the nuclear equation being a TDSE where a time-dependent vector potential and a time-dependent scalar potential (or time-dependent potential energy surface, TDPES)~\cite{Gross_PRL2013, Gross_MP2013, Gross_JCP2015, Suzuki_PRA2014, Suzuki_arXiv2015} represent the effect of the electrons on the nuclei beyond-BO. Therefore, in this context, the electronic equation generates the proper evolution expected when the coupling between electrons and nuclei is fully accounted for and it allows to recover the BO equation in a certain limit, as will be discussed in the paper.

Two major results will be reported: (i) NVPT~\cite{Scherrer_JCTC2013} will be rigorously derived, using as starting point the exact electronic equation from the factorization rather than the complete-adiabatic approach~\cite{Nafie_JCP1983}, and (ii) correction terms to the ``standard'' expression of the rotational strength will naturally appear in the new formulation, due to the presence of the time-dependent vector potential of the theory. Throughout the paper, we will adopt a time-dependent picture, as this is crucial to introduce the concept of nuclear velocity and, thus, to make the connection with NVPT. In such a time-dependent picture we will have access to the instantaneous expectation values of the electric current and of the magnetic dipole moment. The corrections to those expectation values, and therefore to the rotational strength, can be derived also in a static picture referring to the time-independent formulation~\cite{Gross_PTRSA2014, *Gross_arXIv2005} of the factorization,  but the direct link  to NVPT would then be missing.

The paper is organized as follows. In Section~\ref{sec: vcd} we review the linear response theory approach to VCD, showing the connection between rotational strength and intensity of the spectrum. In Section~\ref{sec: factorization} we recall the exact factorization formalism. In Section~\ref{sec: NVPT} we focus on the electronic equation from the exact factorization, showing how to recover the BO limit and introducing non-adiabatic effects as a perturbation to the adiabatic framework. The perturbation parameter is identified as the nuclear velocity, exactly as in NVPT, if the classical limit is considered. However, here we have access to the quantum electronic evolution equation, thus the perturbation parameter has a more general meaning since we are not restricted to a classical treatment of the nuclei. We derive the expressions of the quantities necessary to evaluate the VCD spectrum in Section~\ref{sec: dipoles}, while in Section~\ref{sec: dfpt} we discuss details on the practical calculation of the rotational strength by applying density functional perturbation theory (DFPT). In Section~\ref{sec: examples} we report numerical results for the calculation of rotational strengths and of their corrections for (S)-d$_2$-oxirane, a chiral system, in comparison to oxirane, a non-chiral molecule. We also report the comparison between the NVPT approach and the more standard magnetic field perturbation theory~\cite{Stephens_JPC1985} in Section~\ref{sec: MFPT vs. NVPT}, for (S)-d$_2$-oxirane, (R)-propylene-oxide and (R)-fluoro-oxirane. Our conclusions are stated in Section~\ref{sec: conclusions}.

\section{Theoretical background}\label{sec: theory}

\subsection{Vibrational circular dichroism}\label{sec: vcd}
Vibrational spectroscopy probes the coupling of the nuclear degrees of freedom to applied electro-magnetic fields. On the macroscopic level, the absorption process is described phenomenologically by the Beer-Lambert law~\cite{Barron2004}, where the material specific attenuation of the radiation per unit length is accounted for by the molar absorption coefficient $\epsilon$. Microscopically, and within the linear response regime, the energy dissipated in the interaction between the medium and the radiation is expressed in terms of the observable that couples to the external field. In case of radiation in the infrared spectral range, the multipole approximation and the long wavelength limit can be applied~\cite{Caldwell1971, Barron2004} to determine such coupling. The microscopic and the macroscopic perspectives can be connected in the framework of linear response theory~\cite{Kubo1985}. In the Heisenberg formulation, the frequency dependent absorption coefficient takes the form of the power spectrum of the dipole auto-correlation~\cite{Gordon-1965,McQuarrie1976}.

The specific feature of VCD is the different interaction of chiral systems with polarized light. Linearly polarized light encounters optical rotatory dispersion while circularly polarized light encounters different absorptions for the different handednesses of the radiation, the vibrational circular dichroism (VCD). Formally, this is accounted for by the dependence of the refractive index of a chiral system on the handedness of the radiation. While this effect is not relevant for the mean infrared absorption, the difference absorption gives rise to the VCD signal.

For the calculation of the absorption coefficient $\overline{\epsilon}(\omega)$ and of the difference absorption $\Delta \epsilon(\omega)$~\cite{Nafie}, a common approach in the literature is to invoke the double harmonic approximation for nuclear motion and dipole moment. This leads to the expressions
\begin{align}\label{eqn: ir_harmonic}
 \overline{\epsilon}(\omega) &= \frac{8\pi^3}{3 V h c n(\omega)} \sum_k D_k \omega \delta(\omega - \omega_k)
\end{align}
and
\begin{align}\label{eqn: vcd_harmonic}
 \Delta \epsilon(\omega) &= 4 \frac{8\pi^3}{3 V h c n(\omega)} \sum_k R_k \omega \delta(\omega - \omega_k).
\end{align}
The dipole strength $D_k$ and rotational strength $R_k$ of the vibrational mode $k$, with frequency $\omega_k$, are evaluated as
\begin{align}
 D_k = \frac{\partial \langle \hat{\dot{\bm{\mu}}} \rangle}{\partial\dot q_k} \cdot \frac{\partial \langle \hat{\dot{\bm{\mu}}} \rangle}{\partial \dot q_k} \langle\dot q_k \rangle^2 \label{eqn: dip_str_pos 1}\\
 R_k = \frac{\partial \langle \hat {\bf m} \rangle}{\partial\dot q_k} \cdot \frac{\partial \langle \hat{\dot{\bm \mu}} \rangle}{\partial \dot q_k} \langle \dot q_k \rangle^2, \label{eqn: dip_str_pos 2}
\end{align}
respectively, with the time derivative of the dipole moment $\hat{\dot{\bm{\mu}}}$, namely the current, and the magnetic dipole moment $\hat{\mathbf m}$. In Eqs.~(\ref{eqn: ir_harmonic}) and~(\ref{eqn: vcd_harmonic}), $V$ indicates the volume occupied by the system, $h=2\pi\hbar$ is the Planck constant, $c$ is the speed of light, and $n(\omega)$ is the refractive index of the medium. Normal modes will be indicate throughout the paper as $\mathbf q$, with velocities $\dot{\mathbf q}$.

The linear variations of the expectation values (over the instantaneous state of the system) of the current and of the magnetic dipole moment with respect to (w.r.t.) the mode $q_k$ around their equilibrium values are calculated from the total (electronic and nuclear) atomic polar tensor $\mathcal{P}^\nu$ (APT) and atomic axial tensor $\mathcal{M}^\nu$ (AAT). The APT and AAT have electronic and nuclear contributions~\cite{Nafie,Scherrer_JCTC2013}, namely
\begin{align}
 \frac{\partial \langle \hat{\dot\mu}_\beta \rangle}{\partial\dot R^\nu_\alpha} &\equiv \mathcal{P}^\nu_{\alpha\beta} = \mathcal{E}^\nu_{\alpha\beta} 
\hspace{3pt} + \mathcal{N}^\nu_{\alpha\beta} \\
 \frac{\partial \langle \hat{m}_\beta \rangle}{\partial \dot{R}^\nu_\alpha} &\equiv \mathcal{M}^\nu_{\alpha\beta} = \mathcal{I}^\nu_{\alpha\beta} + \mathcal{J}^\nu_{\alpha\beta},
\end{align}
with electronic parts $\mathcal{E}$ and $\mathcal{I}$ and nuclear parts $\mathcal{N}$ and $\mathcal{J}$. Here, the indices $\alpha$ and $\beta$ are used for the Cartesian coordinates, while $\nu$ labels the nuclei. The dipole and rotational strengths are related via the chain rule to the vibrational nuclear displacement vector $S^\nu_{\alpha k}$ which describes the displacement of nucleus $\nu$ in direction $\alpha$ due to the $k$-th normal mode $q_k$,
\begin{align}\label{eq:naptr}
 S^\nu_{\alpha k}  = \left.\frac{\partial \dot{R}^\nu_\alpha}{\partial \dot{q}_k}\right|_{\dot{{\mathbf q}}=0}
 = \left.\frac{\partial R^\nu_\alpha}{\partial q_k}\right|_{{\mathbf q}=0} .
\end{align}

\subsection{Exact factorization of the electron-nuclear wave function}\label{sec: factorization}
The non-relativistic Hamiltonian describing a system of interacting electrons and nuclei, in the absence of a time-dependent external field, is
\begin{equation}\label{eqn: hamiltonian}
 \hat H = \hat T_n+\hat H_{BO},
\end{equation}
where $\hat T_n$ is the nuclear kinetic energy operator and 
\begin{equation}\label{eqn: boe}
\hat{H}_{BO}(\mathbf r,\mathbf R) = \hat{T}_e(\mathbf r) + \hat{W}_{ee}(\mathbf r) + \hat{V}_{en}(\mathbf r,\mathbf R) +
\hat{W}_{nn}(\mathbf R)
\end{equation}
is the standard BO electronic Hamiltonian, with electronic kinetic energy $\hat{T}_e(\mathbf r)$, and with potentials  $\hat{W}_{ee}(\mathbf r)$  for electron-electron, $\hat{W}_{nn}(\mathbf R)$ for nucleus-nucleus, and  $\hat{V}_{en}(\mathbf r,\mathbf R)$ for electron-nucleus interaction. The symbols $\mathbf r$ and $\mathbf R$ are used to collectively indicate the coordinates of $N_{e}$ electrons and $N_n$ nuclei, respectively.

It has been proved~\cite{Gross_PRL2010,Gross_JCP2012} that the full time-dependent electron-nuclear wave function $\Psi(\mathbf r,\mathbf R,t)$ that is the solution of the TDSE,
\begin{equation}\label{eqn: tdse}
 \hat H\Psi(\mathbf r,\mathbf R,t)=i\hbar\partial_t \Psi(\mathbf r,\mathbf R,t),
\end{equation}
can be exactly factorized to the product
\begin{equation}\label{eqn: factorization}
 \Psi(\mathbf r,\mathbf R,t)=\Phi_{\mathbf R}(\mathbf r,t)\chi(\mathbf R,t)
\end{equation}
 where 
\begin{equation}
 \int d\mathbf r \left|\Phi_{\mathbf R}(\mathbf r,t)\right|^2 = 1 \quad\forall\,\,\mathbf R,t.
\label{eq:PNC}
\end{equation}
Here, $\chi(\mathbf R,t)$ is the nuclear wave function and $\Phi_{\mathbf R}(\mathbf r,t)$ is the electronic wave function  which parametrically depends on the nuclear positions and satisfies the partial normalization condition (PNC) expressed in Eq.~(\ref{eq:PNC}). The PNC guarantees the interpretation of $|\chi(\mathbf R,t)|^2$ as the probability of finding the nuclear configuration $\mathbf R$ at time $t$, and of $|\Phi_{\mathbf R}(\mathbf r,t)|^2$ itself as the conditional probability of finding the electronic configuration $\mathbf r$ at time $t$, given the nuclear configuration $\mathbf R$.  Further, the PNC makes the factorization~(\ref{eqn: factorization}) unique up to within a $(\mathbf R,t)$-dependent gauge transformation,
\begin{equation}\label{eqn: gauge}
 \begin{array}{rcl}
  \chi(\mathbf R,t)\rightarrow\tilde\chi(\mathbf R,t)&=&e^{-\frac{i}{\hbar}\theta(\mathbf R,t)}\chi(\mathbf R,t) \\
  \Phi_{\mathbf R}(\mathbf r,t)\rightarrow\tilde\Phi_{\mathbf R}(\mathbf r,t)&=&e^{\frac{i}{\hbar}\theta(\mathbf R,t)}\Phi_{\mathbf R}(\mathbf r,t).
 \end{array}
\end{equation}
 where $\theta(\mathbf R,t)$ is some real function of the nuclear coordinates and time. 

The stationary variations~\cite{frenkel} of the quantum mechanical action w.r.t. $\Phi_{\mathbf R}(\mathbf r,t)$ and $\chi(\mathbf R,t)$ lead to the equations of motion
\begin{eqnarray}
 \left(\hat H_{el}(\mathbf r,\mathbf R)-\epsilon(\mathbf R,t)\right)
 \Phi_{\mathbf R}(\mathbf r,t)&=&i\hbar\partial_t \Phi_{\mathbf R}(\mathbf r,t)\label{eqn: exact electronic eqn} \\ 
 \hat H_n(\mathbf R,t)\chi(\mathbf R,t)&=&i\hbar\partial_t \chi(\mathbf R,t), \label{eqn: exact nuclear eqn}
\end{eqnarray}
where the PNC is inserted by means of Lagrange multipliers~\cite{Alonso_JCP2013, Gross_JCP2013}. Here, the electronic and nuclear Hamiltonians are defined as
\begin{equation}\label{eqn: electronic hamiltonian}
\hat H_{el}(\mathbf r,\mathbf R)=\hat{H}_{BO}(\mathbf r,\mathbf R)+\hat U_{en}^{coup}[\Phi_{\mathbf R},\chi]
\end{equation}
and
\begin{equation}\label{eqn: nuclear hamiltonian}
\hat H_n(\mathbf R,t) = \sum_{\nu=1}^{N_n} \frac{\left[-i\hbar\nabla_\nu+\bA_\nu(\mathbf R,t)\right]^2}{2M_\nu} + \epsilon(\mathbf R,t),
\end{equation}
respectively, with the ``electron-nuclear coupling operator''
\begin{align}
\hat U_{en}^{coup}&[\Phi_{\mathbf R},\chi]=\sum_{\nu=1}^{N_n}\frac{1}{M_\nu}\Big[
 \frac{\left[-i\hbar\nabla_\nu-\bA_\nu(\mathbf R,t)\right]^2}{2} +\label{eqn: enco} \\
& \left(\frac{-i\hbar\nabla_\nu\chi}{\chi}+\bA_\nu(\mathbf R,t)\right)
 \left(-i\hbar\nabla_\nu-\bA_{\nu}(\mathbf R,t)\right)\Big].\nonumber
\end{align}
The time-dependent potentials are the TDPES, $\epsilon(\mathbf R,t)$, implicitly defined by Eq.~(\ref{eqn: exact electronic eqn}) as
\begin{equation}\label{eqn: tdpes}
 \epsilon(\mathbf R,t)=\left\langle\Phi_{\mathbf R}(t)\right|\hat{H}_{BO}+\hat U_{en}^{coup}-i\hbar\partial_t\left|
 \Phi_{\mathbf R}(t)\right\rangle_{\mathbf r},
\end{equation}
and the vector potential, $\bA_{\nu}\left(\mathbf R,t\right)$, defined as
\begin{equation}\label{eqn: vector potential}
 \bA_{\nu}\left(\mathbf R,t\right) = \left\langle\Phi_{\mathbf R}(t)\right|-i\hbar\nabla_\nu\left.\Phi_{\mathbf R}(t)
 \right\rangle_{\mathbf r}\,.
\end{equation}
The symbol $\left\langle\,\,\cdot\,\,\right\rangle_{\mathbf r}$ indicates an integration over electronic coordinates only. Under the gauge transformation~(\ref{eqn: gauge}), the scalar potential and the vector potential transform as  
\begin{eqnarray}
\tilde{\epsilon}(\mathbf R,t) &=& \epsilon(\mathbf R,t)+\partial_t\theta(\mathbf R,t)\label{eqn: transformation of epsilon} \\
\tilde{\bf A}_{\nu}(\mathbf R,t) &=& {\bf A}_{\nu}(\mathbf R,t)+\nabla_\nu\theta(\mathbf R,t)\,.\label{eqn: transformation of A}
\end{eqnarray}
In Eqs.~(\ref{eqn: exact electronic eqn}) and~(\ref{eqn: exact nuclear eqn}), $\hat U_{en}^{coup}[\Phi_{\mathbf R},\chi]$, $\epsilon(\mathbf R,t)$, and $\bA_{\nu}\left(\mathbf R,t\right)$ are responsible for the coupling between electrons and nuclei in a formally exact way. It is worth noting that the electron-nuclear coupling operator, $\hat U_{en}^{coup}[\Phi_{\mathbf R},\chi]$ in the electronic equation~(\ref{eqn: exact electronic eqn}) depends on the nuclear wave function and acts on the parametric dependence of $\Phi_{\mathbf R}(\mathbf r,t)$ as a differential operator. This ``pseudo-operator'' includes the coupling to the nuclear subsystem beyond the parametric dependence in the BO Hamiltonian $\hat H_{BO}(\mathbf r,\mathbf R)$.

The nuclear equation~(\ref{eqn: exact nuclear eqn}) has the particularly appealing form of a Schr\"odinger equation that contains the TDPES~(\ref{eqn: tdpes}) and the vector potential~(\ref{eqn: vector potential}) governing nuclear dynamics and yielding the nuclear wave function. The scalar and vector potentials are uniquely determined up to within a gauge transformation, given by Eqs.~(\ref{eqn: transformation of epsilon}) and~(\ref{eqn: transformation of A}). As expected, the nuclear Hamiltonian in Eq.~(\ref{eqn: exact nuclear eqn}) is form-invariant under such transformations. $\chi(\mathbf R,t)$ is interpreted as the nuclear wave function since it leads to an $N$-body nuclear density and an $N$-body current-density which reproduce the true nuclear $N$-body density and current-density~\cite{Gross_JCP2012} obtained from the full wave function $\Psi(\mathbf r,\mathbf R,t)$. The uniqueness of $\epsilon(\mathbf R,t)$ and $\bA_{\nu}(\mathbf R,t)$ can be straightforwardly proved by following the steps of the current-density version~\cite{Dhara_PRA1988} of the Runge-Gross theorem~\cite{Gross_PRL1984}, or by referring to the theorems proved in Ref.~\cite{Gross_PRL2010}. 

\section{Nuclear velocity perturbation theory}\label{sec: NVPT}
Before showing the derivation of the velocity-dependent corrections to the BO wave function within the exact factorization approach, let us present a procedure that allows us to recover the BO limit of the electronic equation~(\ref{eqn: exact electronic eqn}). Suppose first that the electron-nuclear wave function is given as a BO product
\begin{align}
\Psi\left(\mathbf r,\mathbf R,t\right)=\Phi_{\mathbf R}(\mathbf r,t) \chi\left(\mathbf R,t\right)= \varphi_{\mathbf R}^{(0)}\left(\mathbf 
r\right)\chi\left(\mathbf R,t\right).
\end{align}
Here, $\varphi_{\mathbf R}^{(0)}\left(\mathbf r\right)$ indicates the real and not degenerate BO ground-state. Using Eq.~(\ref{eqn: vector potential}), it follows that the vector potential vanishes identically, $\bA_\nu(\mathbf R,t)\equiv0$~\cite{Min_PRL2014}. This can be interpreted as a choice of gauge~\footnote{This choice of gauge is in general not possible. For instance, when the electronic state is degenerate or when the electronic wave function has singularities in the parameter space, $\mathbf R$, $\varphi_{\mathbf R}^{(0)}\left(\mathbf r\right)$ cannot be real and thus the vector potential might be non-zero.}. With this assumption, the electron-nuclear coupling operator from Eq.~(\ref{eqn: enco}) becomes
\begin{equation}\label{eqn: enco without A}
\hat U_{en}^{coup}[\Phi_{\mathbf R},\chi]=\sum_{\nu=1}^{N_n}\frac{-\hbar^2\nabla_{\nu}^2}{2M_\nu}+
\frac{-i\hbar\nabla_\nu\chi(\mathbf R,t)}{M_\nu\chi(\mathbf R,t)}\cdot\left(-i\hbar\nabla_\nu\right).
\end{equation}
The first term on the right hand side (RHS), containing the Laplacian~\cite{handyjcp1986, valeevjcp2003, tajtijcp2007} w.r.t. nuclear coordinates, will be neglected from now on. It can be shown, as reported in Refs.~\cite{Gross_EPL2014, Gross_JCP2014, Gross_PRL2015}, that this term contributes with second-order non-adiabatic couplings to the electronic equation, but being explicitly $\mathcal O(M_\nu^{-1})$ its effect can be neglected if compared to the remaining (and leading) term. Following again Refs.~\cite{Gross_EPL2014, Gross_JCP2014, Gross_PRL2015}, the term that depends on $\chi$ can be approximated to zero-th order in an $\hbar$-expansion~\cite{Van-Vleck_PNAS1928} of the nuclear wave function as the classical nuclear velocity, namely
\begin{equation}\label{eqn: classical velocity in enco}
\frac{1}{M_\nu}\frac{-i\hbar\nabla_\nu\chi(\mathbf R,t)}{\chi(\mathbf R,t)} = \frac{\bP_\nu(\mathbf R,t)}{M_\nu}=\dot{\mathbf R}_{\nu}(t).
\end{equation}
We have invoked here the classical limit in order to directly relate our results to the NVPT~\cite{Scherrer_JCTC2013} and to justify the condition of ``small nuclear velocity'' that allows a treatment of effects beyond-BO within perturbation theory. The procedure, however, does not rely on the classical limit and the ``small'' perturbation parameter will be denoted as
\begin{align}\label{eqn: parameter}
\boldsymbol{\lambda}_\nu(\mathbf R,t) = \frac{1}{M_\nu}\frac{-i\hbar\nabla_\nu\chi(\mathbf R,t)}{\chi(\mathbf R,t)}.
\end{align}
Eq.~(\ref{eqn: parameter}) contains the variations in space of the phase and of the modulus of the nuclear wave function~\cite{semiclassics}, and when both variations are ``small'' then the approach considered here can be applied. We have justified the former hypothesis (small variations of the phase) by employing the classical approximation and we are now assuming valid also the latter (small variations of the modulus).

The electronic Hamiltonian from Eq.~(\ref{eqn: electronic hamiltonian}) becomes
\begin{equation}\label{eqn: el eqn 1}
\hat H_{el}(\mathbf r,\mathbf R) = \hat H_{BO}+\sum_{\nu=1}^{N_n} \boldsymbol{\lambda}_\nu(\mathbf R,t)\cdot \left(-i\hbar\nabla_\nu\right)
\end{equation}
and the TDPES reads
\begin{align}\label{eqn: BO as TDPES}
\epsilon(\mathbf R,t) &= \left\langle\varphi_{\mathbf R}^{(0)}\right|\hat H_{BO}+\sum_{\nu=1}^{N_n} \boldsymbol{\lambda}_\nu(\mathbf R,t)\cdot \left(-i\hbar\nabla_\nu\right)\left|\varphi_{\mathbf R}^{(0)}\right\rangle_{\mathbf r} \nonumber \\
&=\epsilon_{BO}^{(0)}(\mathbf R),
\end{align}
i.e. only the $\hat H_{BO}$ term survives, since the second term does not contribute to the TDPES. Notice that here the term $\langle\Phi_{\mathbf R}(t)|-i\hbar\partial_t|\Phi_{\mathbf R}(t)\rangle_{\mathbf r}$ identically vanishes, because the electronic wave function is the time-independent BO wave function. In order to recover from Eq.~(\ref{eqn: el eqn 1}) the electronic equation within the BO approximation, one should impose $\boldsymbol{\lambda}_\nu(\mathbf R,t)=0$, or similarly $\dot{\mathbf R}_{\nu}(t)=0$ $\forall\,\nu$ as the electronic equation in BO is solved for fixed nuclei (meaning that their velocity is zero).

To summarize, in order to construct the Hamiltonian in Eq.~(\ref{eqn: el eqn 1}) (i) we treat the nuclei classically, thus we consider the nuclear wave function up to within $\mathcal O(\hbar^0)$ terms, (ii) we derive corrections to the BO Hamiltonian that are proportional to the nuclear velocity, thus recovering the BO electronic equation if the nuclear velocity is zero (condition of fixed nuclei), (iii) we ``relax'' the hypothesis of classical nuclei by introducing $\boldsymbol{\lambda}_\nu(\mathbf R,t)$ as the perturbation parameter.

Combining Eqs.~(\ref{eqn: el eqn 1}) and~(\ref{eqn: BO as TDPES}) will provide the electronic equation within the new formulation of NVPT. In contrast to the formulation based on the complete-adiabatic approach~\cite{Nafie_JCP1983}, the perturbative scheme presented here directly applies to the electronic equation rather than to the full TDSE. Using perturbation theory~\cite{Scherrer_JCTC2013}, where $\hat H_{BO}$ is the unperturbed Hamiltonian and the second term on the RHS of Eq.~(\ref{eqn: el eqn 1}) is the perturbation, we find the solutions of the equation
\begin{equation}\label{eqn: el eqn in NVPT}
\left[\hat H_{BO}-\epsilon^{(1)}-i\hbar\sum_{\nu=1}^{N_n} \boldsymbol{\lambda}_\nu(\mathbf R,t)\cdot \nabla_\nu\right]\varphi_{\mathbf R}(\mathbf r,t) = 0,
\end{equation}
as
\begin{align}
\varphi_{\mathbf R}(\mathbf r,t) &= \varphi_{\mathbf R}^{(0)}(\mathbf r)\nonumber \\
 &+ \sum_{e\neq 0}\frac{\left\langle\varphi_{\mathbf R}^{(e)}\left|-i\hbar\sum_{\nu,\alpha}
\lambda^{\nu}_\alpha(\mathbf R,t)\partial_{\alpha}^{\nu}\right.\!\varphi_{\mathbf R}^{(0)}\right\rangle_{\mathbf r}}{\epsilon_{BO}^{(0)}
\left(\mathbf R\right)-\epsilon_{BO}^{(e)}\left(\mathbf R\right)}\varphi_{\mathbf R}^{(e)}\left(\mathbf r\right)
\label{eqn: 0th and 1st order}
\end{align}
up to within linear-order in the perturbation, with the index $\nu$ running over the $N_n$ nuclei and with $\alpha$ running over the three Cartesian coordinates. The symbol $\partial_\alpha^\nu$ is used to indicate a spatial derivative along the $\alpha$ direction of the position of the $\nu$-th nucleus and $e$ labels the (unperturbed) adiabatic excited states. The TDPES, up to within first-order terms, is labeled $\epsilon^{(1)}$. It is worth noting that in writing Eq.~(\ref{eqn: el eqn in NVPT}), we are discarding the variations in time of the first-order correction to the BO wave function, adopting a previously assumed~\cite{Scherrer_JCTC2013} hypothesis that these variations are smaller than the perturbation itself, thus negligible at the given order. We re-write Eq.~(\ref{eqn: 0th and 1st order}) as
\begin{equation}\label{eqn: perturbed electronic wf}
\varphi_{\mathbf R}(\mathbf r,t)=\varphi_{\mathbf R}^{(0)}(\mathbf r)+\sum_{\nu,\alpha} i\lambda_{\alpha}^\nu(\mathbf R,t)\varphi_{\mathbf R,\nu\alpha}^{(1)}(\mathbf r),
\end{equation}
introducing the definition of the first-order perturbation to the BO ground-state
\begin{align}
\varphi_{\mathbf R,\nu\alpha}^{(1)}(\mathbf r) = \sum_{e\neq 0}\frac{d_{e0,\nu\alpha}\left(\mathbf R\right)}{\omega_{e0}(\mathbf R)}
\varphi_{\mathbf R}^{(e)}\left(\mathbf r\right).\label{eqn: perturbation with NACV}
\end{align}
Here $d_{e0,\nu\alpha}\left(\mathbf R\right)$ is the $\alpha$-th Cartesian component of the non-adiabatic coupling vector, corresponding to the $\nu$-th nucleus, between the unperturbed ground-state and the excited state $e$, whereas the frequency $\omega_{e0}(\mathbf R)$ is the the energy difference (divided by $\hbar$) between the excited ($e$) and the ground ($0$) states. When the adiabatic states are real, Eq.~(\ref{eqn: perturbation with NACV}) is real as well and the second term in Eq.~(\ref{eqn: perturbed electronic wf}) is purely imaginary. Moreover, the correction term in Eq.~(\ref{eqn: perturbed electronic wf}) depends on time only implicitly, via its dependence on $\boldsymbol{\lambda}_\nu(\mathbf R,t)$, and $\varphi_{\mathbf R,\nu\alpha}^{(1)}(\mathbf r)$ is orthogonal to $\varphi_{\mathbf R}^{(0)}(\mathbf r)$. This last property can be interpreted as a choice of gauge. For instance, by imposing the condition that $\langle \varphi_{\mathbf R}^{(0)}|\varphi_{\mathbf R}(t)\rangle_{\mathbf r}$ is real $\forall\,\mathbf R,t$, which is allowed as gauge condition, we imply the orthogonality of $\varphi_{\mathbf R,\nu\alpha}^{(1)}(\mathbf r)$ and $\varphi_{\mathbf R}^{(0)}(\mathbf r)$, namely $\langle \varphi_{\mathbf R}^{(0)}|\varphi_{\mathbf R,\nu\alpha}^{(1)}\rangle_{\mathbf r}=0$. It easy to prove that the PNC remains valid up to within $\mathcal O(\lambda_\alpha^\nu)$, using the orthogonality of $\varphi_{\mathbf R}^{(0)}(\mathbf r)$ and $\varphi_{\mathbf R,\nu\alpha}^{(1)}(\mathbf 
r)$.

The first-order approximation to the TDPES is
\begin{align}
\epsilon^{(1)}(\mathbf R,t) &= \epsilon_{BO}^{(0)}\left(\mathbf R\right)-i\sum_{\nu}\mathcal O(\bm{\lambda}_\nu(\mathbf R,t))
\label{eqn: 0th and 1st order energy}
\end{align}
but the second term on the RHS is identically zero, as can be proved by either inserting Eq.~(\ref{eqn: perturbed electronic wf}) in the definition of the TDPES given in Eq.~(\ref{eqn: tdpes}) or by considering the fact that $\epsilon^{(1)}(\mathbf R,t)$ must be real while the correction is purely imaginary.

As in the NVPT approach based on the complete-adiabatic form of the electron-nuclear wave function~\cite{Nafie_JCP1983}, the first-order perturbation to the electronic wave function represents the effect of the non-adiabatic coupling between the ground and the excited electronic states. Within a fully non-adiabatic approach~\cite{Tully_JCP1990, Marx_PRL2002, Tavernelli_PRL2002, Sugino_JCP2007, Mukamel_JCP2000, Baer_CPL2002, Tavernelli_JCP2009_1, Tavernelli_JCP2009_2, Tavernelli_JCP2010, Subotnik_JCP2014}, it would be possible to compute Eq.~(\ref{eqn: perturbation with NACV}). However, it has been shown in Ref.~\cite{Scherrer_JCTC2013} that within DFPT the perturbation can be determined by the knowledge of only ground-state properties. Eq.~(\ref{eqn: el eqn in NVPT}) is solved by inserting the chosen expression for the electronic wave function~(\ref{eqn: perturbed electronic wf}) and by solving for each order in the perturbation $\boldsymbol{\lambda}_\nu(\mathbf R,t)$. At the zero-th order we obtain
\begin{equation}\label{eqn: zeroth order in velocity}
\Big[\hat H_{BO}-\epsilon_{BO}^{(0)}(\mathbf R)\Big] \varphi_{\mathbf R}^{(0)}(\mathbf r) = 0
\end{equation}
and at the first-order
\begin{align}\label{eqn: first order in velocity}
\Big[\hat H_{BO}-\epsilon_{BO}^{(0)}(\mathbf R)\Big]\varphi_{\mathbf R,\nu\alpha}^{(1)}(\mathbf r) =\hbar\partial_\alpha^\nu\varphi_{\mathbf R}^{(0)}(\mathbf r) \,\,\forall\,\nu,\alpha.
\end{align}
Eq.~(\ref{eqn: zeroth order in velocity}) is simply the eigenvalue problem associated to the BO Hamiltonian; Eq.~(\ref{eqn: first order in velocity}) is solved in the framework of DFPT as illustrated in the Section~\ref{sec: dfpt}.

The TDPES of the theory based on the exact factorization remains unaffected if compared to the BO case, up to within the first-order perturbation, as shown in Eq.~(\ref{eqn: 0th and 1st order energy}). The vector potential, that is identically zero in the adiabatic treatment, becomes
\begin{align}
\bA_{\nu}(\mathbf R,t) =-2\hbar\sum_{\nu',\alpha}\lambda_\alpha^{\nu'}(\mathbf R,t)\left\langle \nabla_{\nu}\varphi_{\mathbf 
R}^{(0)}\right.\left|\varphi_{\mathbf R,\nu'\alpha}^{(1)}\right\rangle_{\mathbf r}\label{eqn: first expression of A}
\end{align}
This expression is obtained by using Eq.~(\ref{eqn: perturbed electronic wf}) in the definition of the vector potential given in Eq.~(\ref{eqn: vector potential}). Using Eq.~(\ref{eqn: first order in velocity}) in Eq.~(\ref{eqn: first expression of A}), an alternative expression is derived, that is used in actual calculations, namely
\begin{align}
A_\alpha^\nu&(\mathbf R,t) =-\sum_{\nu',\beta}\lambda_\beta^{\nu'}(\mathbf R,t) \mathcal A^{\nu\nu'}_{\alpha\beta}\left(\mathbf R\right),\label{eqn: third expression of A}\\
=&-2 \sum_{\nu',\beta}\lambda_\beta^{\nu'}(\mathbf R,t)\left\langle\varphi_{\mathbf R,\nu'\beta}^{(1)} \right|\hat 
H_{BO}-\epsilon_{BO}(\mathbf R)\left|\varphi_{\mathbf R,\nu\alpha}^{(1)}\right\rangle_{\mathbf r}\label{eqn: second expression of A} 
\end{align}
where we have introduced the definition of the matrix $\mathcal A^{\nu\nu'}_{\alpha\beta}\left(\mathbf R\right)$ and the symbol $A_\alpha^\nu$ stands for the $\alpha$ Cartesian coordinate of the vector potential corresponding to the $\nu$-th nucleus. It is instructive to give an alternative formula for the evaluation of the vector potential matrix in Eq.~(\ref{eqn: third expression of A}), namely
\begin{align}
\mathcal A^{\nu\nu'}_{\alpha\beta}\left(\mathbf R\right) = 2\hbar\sum_{e\neq 0}\frac{d_{e0,\nu'\beta}(\mathbf R)d_{e0,\nu\alpha}(\mathbf R)}{\omega_{e0}(\mathbf R)},
\end{align}
that is obtained by using Eq.~(\ref{eqn: perturbation with NACV}) in Eq.~(\ref{eqn: second expression of A}) and acting with the BO Hamiltonian on its eigenstates. This expression is useful to determine the vector potential by combining the NVPT with (explicit) non-adiabatic calculations. In general, evaluating the vector potential from the full electronic wave function in Eq.~(\ref{eqn: factorization}) is difficult because the exact electronic state is not known, thus approximations have to be invoked. Here, we have derived an expression that can instead be used in actual calculations. However, in the present paper we focus on Eq.~(\ref{eqn: second expression of A}) and we estimate it within DFPT. 

\section{Observables}\label{sec: observables}
\subsection{Current and magnetic dipole moment}\label{sec: dipoles}
In a time-dependent picture, the expectation values of the current and of the magnetic dipole moment on the instantaneous state of the system are employed to evaluate the rotational strength giving access to the VCD spectrum in the linear response regime. We will derive their expressions employing the factorized form of the full wave function when calculating explicitly the expectation values.

The current and magnetic dipole moment operators are defined as 
\begin{equation}\label{eqn: electric dipole 1}
\hat{\dot{\boldsymbol\mu}} = \hat{\dot{\boldsymbol\mu}}^e+\hat{\dot{\boldsymbol\mu}}^n=
-\sum_{i=1}^{N_e}\frac{e}{m}\hat{\mathbf p}_i+\sum_{\nu=1}^{N_n}\frac{Z_{\nu}e}{M_\nu}\hat{\mathbf P}_\nu,
\end{equation}
and
\begin{equation}\label{eqn: magnetic dipole 1}
{\hat{\bf m}} = \hat{\bf m}^e+\hat{\bf m}^n=
-\sum_{i=1}^{N_e}\frac{e}{2mc}\hat{\mathbf r}_i\times \hat{\mathbf p}_i+\sum_{\nu=1}^{N_n}\frac{Z_{\nu}e}{2M_{\nu}c}\hat{\mathbf R}_\nu \times 
\hat{\mathbf P}_\nu,
\end{equation}
respectively. Here, $e$ is the electronic charge, $Z_\nu e$ is the nuclear charge, $m$ and $M_\nu$ are the electronic and nuclear masses and $c$ is the speed of light. The position and momentum operators for the electronic subsystem are indicated as $\hat{\mathbf r}_i$ and $\hat{\mathbf p}_i$, respectively, and similar symbols are used for the nuclear operators, $\hat{\mathbf R}_\nu$ and $\hat{\mathbf P}_\nu$. As expected, the vector potential does not appear in Eqs.~(\ref{eqn: electric dipole 1}) and~(\ref{eqn: magnetic dipole 1}) since we are not yet calculating an expectation value. However, since the nuclear momentum operator in position representation acts as a derivative w.r.t. the nuclear coordinates $\mathbf R$, the vector potential appears (only) when the derivative acts on the parametric dependence of the electronic wave function. Indeed, if the factorization is not introduced, such vector potential will never be present.

The expectation values of the operators in Eqs.~(\ref{eqn: electric dipole 1}) and~(\ref{eqn: magnetic dipole 1}) on $\Psi(\mathbf r,\mathbf R,t)$ are indicated with the symbol $\langle\,\cdot\,\rangle_{\Psi}$,
\begin{align}\label{eqn: current on psi}
\left\langle{\hat{\dot{\boldsymbol\mu}}}\right\rangle_{\Psi} =\int d\mathbf R \chi^*(\mathbf R,&t)\Big[ \left\langle\Phi_{\mathbf R}(t)\left|\hat{\dot{\boldsymbol\mu}}^e\right|\Phi_{\mathbf R}(t)\right\rangle_{\mathbf r}\nonumber\\
&+\hat{\dot{\boldsymbol\mu}}^n+\sum_{\nu=1}^{N_n}\frac{Z_\nu e}{M_\nu}\mathbf A_\nu\left(\mathbf R,t\right)\Big]\chi(\mathbf R,t)
\end{align}
and
\begin{align}\label{eqn: magnetic dipole on psi}
\left\langle{\hat{\bf m}}\right\rangle_{\Psi} =&
\int d\mathbf R \chi^*(\mathbf R,t)\Bigg[\left\langle\Phi_{\mathbf R}(t)\left|\hat{\bf m}^e\right|\Phi_{\mathbf R}(t)\right\rangle_{\mathbf 
r}\nonumber \\
&+\hat{\bf m}^n+\sum_{\nu=1}^{N_n}\frac{Z_\nu e}{2M_\nu c}\hat{\mathbf R}_{\nu} \times \mathbf A_{\nu}(\mathbf R,t)\Bigg]\chi(\mathbf R,t).
\end{align}
We will now introduce the following symbols for the expectation values of the electronic contributions to the current and magnetic dipole moment on the (exact) electronic wave function,
\begin{align}
\dot{\boldsymbol\mu}^e_{\mathbf R}(t) &= \left\langle\Phi_{\mathbf R}(t)\left|\hat{\dot{\boldsymbol\mu}}^e\right|\Phi_{\mathbf R}(t)\right\rangle_{\mathbf r} \\
\mathbf m_{\mathbf R}^e(t) & =\left\langle\Phi_{\mathbf R}(t)\left|\hat{\bf m}^e\right|\Phi_{\mathbf R}(t)\right\rangle_{\mathbf r}.
\end{align}
If the BO electronic wave function is used to approximate $\Phi_{\mathbf R}(\mathbf r,t)$, both equations, i.e. the electronic contributions to the expectation values, vanish, as well as the vector potential in Eqs.~(\ref{eqn: current on psi}) and~(\ref{eqn: magnetic dipole on psi}), as mentioned above. It is, however, now possible to insert the NVPT approximation to the electronic wave function, Eq. (\ref{eqn: perturbed electronic wf}), and this leads to the following expressions for the expectation values,
\begin{align}
\left\langle{\hat{\dot{\boldsymbol\mu}}}\right\rangle_{\Psi}&\simeq\left\langle {\dot{\boldsymbol\mu}}^{e,(1)}_{\mathbf R}(t)\right\rangle_{\chi} + \sum_{\nu=1}^{N_n} \frac{Z_\nu e}{M_\nu} \left\langle\hat{\mathbf P}_\nu+\mathbf A_{\nu}(\mathbf R,t)\right\rangle_\chi\label{eqn: electric dipole} \\
\left\langle{\hat{\bf m}}\right\rangle_{\Psi} &\simeq\left\langle{\bf m}^{e,(1)}_{\mathbf R}(t)\right\rangle_{\chi}\nonumber \\
&+\sum_{\nu=1}^{N_n}\frac{Z_\nu e}{2M_\nu c}\left\langle\hat{\mathbf R}_{\nu} \times \left[\hat{\mathbf P}_\nu+\mathbf A_{\nu}(\mathbf R,t)\right]\right\rangle_\chi.\label{eqn: magnetic dipole}
\end{align}
Here, we have written the expectation values (on the left hand sides) on $\Psi$, the full electron-nuclear wave function, in terms of expectation values of new observables on $\chi$, the nuclear wave function only. Therefore, the vector potential naturally appears in the equations. In addition, since the electronic wave function has been approximated, as stated above, by using Eq.~(\ref{eqn: perturbed electronic wf}), we obtain that the current and the magnetic dipole moment contain an electronic contribution that is first-order $(1)$ in the perturbation. The second terms in both equations, containing the vector potential, corrects the nuclear contribution to both expectation values and these corrections shall be considered within NVPT since they are first-order in the perturbation parameter $\boldsymbol\lambda_\nu(\mathbf R,t)$ (see Eq.~(\ref{eqn: third expression of A})). Standard approaches do not consider these correction terms, because the vector potential is a quantity that has been introduced only in the context of the exact factorization. We will compute explicitly these corrections in Section~\ref{sec: results}, but we can already anticipate that while the first (standard) term is $\mathcal O(\boldsymbol\lambda_\nu)$, because of the $\hat{\mathbf P}_\nu/M_\nu$ term, the correction is $\mathcal O(\boldsymbol\lambda_\nu/M_\nu)$ since the vector potential itself has a linear dependence on the perturbation parameter.

It is worth mentioning here that the advantage of introducing expectation values on the nuclear wave function, rather than on the full wave function, becomes clear when the classical approximation for the nuclear subsystem is considered. In this case, due to the properties of the nuclear wave function in the factorization framework ($\chi$ is a proper wave function, as it evolves according to a TDSE, and leads to the density and current-density calculated from the full wave function), the classical limit can be performed by imposing that the nuclear density infinitely localizes, at each time, at the classical position denoted by the trajectory. The second terms on the RHS of Eqs.~(\ref{eqn: electric dipole}) and~(\ref{eqn: magnetic dipole}) then become simply functions of phase-space variables. It is important to notice, however, that the vector potential has to be taken into account to appropriately relate the nuclear velocity and momentum.

\subsection{Rotational strengths from density functional perturbation theory}\label{sec: dfpt}
The direct numerical solution of Eqs.~(\ref{eqn: zeroth order in velocity}) and~(\ref{eqn: first order in velocity}) is very expansive for systems with more than a few degrees of freedom. Already the calculation for small chiral molecules requires an approximate treatment of the electronic structure problem. In our implementation we resort to standard Kohn-Sham (KS) DFT~\cite{Hohenberg-1964, Kohn-1965, Jones-1989} with generalized gradient approximation to the exchange-correlation functional~\cite{BLYP_B, BLYP_LYP}. For simplicity, we will limit our discussion to the case of spin saturated closed shell systems and drop the explicit notation of the parametric dependence on the nuclear positions.

In the framework of single determinant KS-DFT, Eq.~(\ref{eqn: zeroth order in velocity}) directly translates to the standard BO ground-state electronic structure problem
\begin{align}
 \Big[\hat H_{KS}^{(0)}-\epsilon_{o}^{(0)}\Big]\phi_{ o}^{(0)}(\mathbf r) = 0
\end{align}
with the unperturbed KS Hamiltonian $\hat H_{KS}^{(0)}$ and the unperturbed KS orbitals $\phi_{o}^{(0)}$ and KS energies $\epsilon_{o}^{(0)}$ of the occupied electronic states $o$. In DFPT~\cite{Gonze-1995a, Gonze-1995b, Putrino-2000, Baroni-2001, Watermann-2014}, the calculation of the non-adiabatic correction to the ground-state orbitals can be done without explicit knowledge of the unoccupied states via a Sternheimer equation~\cite{Sternheimer-1954}
\begin{align}\label{eqn: DFPT general}
 -\hat P_e\Big[\hat H_{KS}^{(0)}-\epsilon_{o}^{(0)}\Big]\hat P_e\phi_{o}^{(1)}(\mathbf r) = \hat P_e \hat H_{KS}^{(1)}[\{\phi_{ o}\}]\phi_{ 
o}^{(0)}(\mathbf r)
\end{align}
with a projector on the manifold of the unoccupied states $\hat P_e=1-\sum_o |\phi_{o}\rangle \langle \phi_{o}|$. The perturbation Hamiltonian on the RHS $\hat H_{KS}^{(1)}[\{\phi_{o}\}]$ can depend on the electronic density response and hence implicitly on the perturbed orbitals on the left hand side. This is the case for electric field or nuclear displacement perturbations and requires a self-consistent solution. Explicitly, Eq.\ (\ref{eqn: DFPT general}) for a nuclear displacement perturbation $j$ reads
\begin{align}\label{eqn: DFPT nuclear displacement perturbation}
 -\hat P_e \Big[\hat H_{KS}^{(0)}-\epsilon_{o}^{(0)}\Big]\hat P_e\frac{\partial \phi_{o}^{(0)}(\mathbf r)}{\partial R_j} =\hat P_e\frac{\partial \hat 
H_{KS}}{\partial R_j}[\{\phi_{ o}\}]\phi_{o}^{(0)}(\mathbf r).
\end{align}
The perturbed KS orbitals $\partial_{R_j} \phi_{o}^{(0)}(\mathbf r)$ are the gradient of the KS orbitals $\phi_{o}^{(0)}(\mathbf r)$ w.r.t. a nuclear displacement $j$. They can be used for the calculation of the electronic APT in the position form~\cite{Nafie, Scherrer_JCTC2013}.

The corresponding translation of Eq.\ (\ref{eqn: first order in velocity}) to DFPT reads
\begin{align}\label{eqn: DFPT nuclear velocity perturbation}
\hat P_e\Big[\hat H_{KS}^{(0)}-\epsilon_{o}^{(0)}\Big]\hat P_e\phi_{o,j}^{(1)}(\mathbf r) =\hat P_e\hbar\partial_{R_j}\phi_{o}^{(0)}(\mathbf r) 
\,\,\forall\,j.
\end{align}
Also this equation is reminiscent of a Sternheimer equation. However, instead of an explicit perturbation Hamiltonian acting on the unperturbed KS orbitals, the RHS is proportional to the gradient of the ground-state wave function w.r.t. a nuclear displacement. As already discussed, this gradient is accessible via Eq.\ (\ref{eqn: DFPT nuclear displacement perturbation}). This method requires two response calculations, a self-consistent one for the nuclear displacement perturbation and another for the nuclear velocity perturbation.

Recently, a related approach to the calculation of NVPT has been reported~\cite{Patchkovskii_JCP2012} which relies on an iterative finite-differences scheme for the construction of the intermediate nuclear gradient information.

With the imaginary correction to the BO electronic wave function in Eq.~(\ref{eqn: perturbed electronic wf}), it is possible to calculate the electronic APT $\mathcal{E}$ in the velocity form
\begin{align}
 \mathcal{E}^\nu_{\alpha\beta} = \frac{\partial \langle \hat{\dot{\mu}}_\beta^e \rangle}{\partial \dot{R}^\nu_\alpha} = 2 \sum_o \langle \phi_o | 
\hat{\dot{\mu}}_\beta^e | \phi^{(1)}_{o,(\nu,\alpha)} \rangle
\end{align}
and the electronic AAT $\mathcal{I}$
\begin{align}
 \mathcal{I}^\nu_{\alpha\beta} = \frac{\partial \langle \hat{m}_\beta^e \rangle}{\partial \dot{R}^\nu_\alpha} = 2 \sum_o \langle \phi_o | \hat m_\beta^e 
| \phi^{(1)}_{o,(\nu,\alpha)} \rangle
\end{align}
For the calculation of the magnetic moment, a choice of the origin of the position operator has to be made. This poses additional complications for the calculation of observables in the condensed phase where periodic boundary conditions are used. For a detailed discussion, we refer to the literature~\cite{Nafie,Scherrer_JCTC2013} and to Appendix~\ref{app: origin dependence}.

The nuclear AAT $\mathcal{J}$ is decomposed into its ``conventional'' contribution and the correction due to the presence of the vector potential
\begin{align}
 \mathcal{J}_{\alpha\beta}^\nu 
 = \frac{Z_\nu e}{2 c} \epsilon_{\alpha\beta\gamma} R^\nu_\gamma + \Delta \mathcal{J}_{\alpha\beta}^\nu
\end{align}
where we used Einstein's summation convention for repeated indices. The correction due to the additional term in the nuclear magnetic moment in Eq.~(\ref{eqn: magnetic dipole}) is given by the derivative of
\begin{align}\label{eqn: nucl mag moment correction vel dep}
 \langle \Delta m_\beta^n \rangle = \frac{Z_{\nu} e}{2 M_{\nu} c} \epsilon_{\beta\gamma\delta} R^{\nu}_\gamma A^{\nu}_\delta = \frac{Z_{\nu} e}{2 
M_{\nu} c} \epsilon_{\beta\gamma\delta} R^{\nu}_\gamma \mathcal A^{\nu\nu'}_{\delta \eta} \lambda^{\nu'}_\eta
\end{align}
w.r.t. $\dot R_\alpha^{\nu}$. Written in this form, the correction to the magnetic moment depends linearly on the nuclear velocities, via the identification $\lambda^{\nu'}_\eta=\dot R^{\nu'}_\eta$. However, this dependence can be removed in the picture of the nuclear AAT. To see this, we evaluate the vector potential matrix of Eq.\ (\ref{eqn: second expression of A}) as
\begin{align}
\mathcal A^{\nu'\nu}_{\delta \eta} = 2 \sum_o \langle \phi^{(1)}_{o,(\nu',\delta)} | \hat H_{KS}^{(0)}-\epsilon_{o}^{(0)} | \phi^{(1)}_{o,(\nu,\eta)} 
\rangle
\end{align}
and take the derivative of Eq.\ (\ref{eqn: nucl mag moment correction vel dep}) w.r.t. a nuclear velocity. This gives the correction to the nuclear AAT as
\begin{align}\label{eqn: nuclear AAT correction second form}
 \Delta \mathcal J_{\alpha\beta}^\nu = \frac{Z_{\nu'} e}{2 M_{\nu'} c} \epsilon_{\beta\gamma\delta} R^{\nu'}_\gamma \mathcal A^{\nu'\nu}_{\delta\alpha}
\end{align}
This expression illustrates two features of the correction. First, it is non-local in the nuclear contributions, i.e. all nuclei contribute to the AAT of a single nucleus. Second, the pre-factor contains the inverse nuclear mass, while the conventional contribution does not.

\section{Numerical results}\label{sec: results}
The presented NVPT has been implemented in our development version of the CPMD~\cite{CPMD-3.15.3, Scherrer_JCTC2013} electronic structure package. The calculations have been performed using DFPT~\cite{Putrino-2000, Baroni-2001, Watermann-2014} with Troullier-Martins~\cite{Troullier-1991} pseudo-potentials and the BLYP~\cite{BLYP_B, BLYP_LYP} functional. We have employed a plane wave cutoff of 100 Ry. The fluorine pseudo-potential with a radius $r_c=1.2$ has been used. The geometry optimizations, harmonic analysis and magnetic field perturbation~\cite{Stephens_JPC1985} calculations were done using the electronic structure program Gaussian 09 Revision D.01~\cite{g09} employing aug-cc-pVTZ basis set~\cite{Kendall1992} and BLYP functional.

\subsection{(S)-d$_2$-oxirane vs. Oxirane}\label{sec: examples}
\begin{figure}[ht!]
 \centering
 \includegraphics*[width=0.3\linewidth]{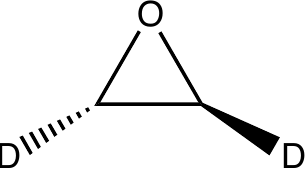}
 \caption{(S)-d$_2$-oxirane.}
 \label{fig:S-d2-oxirane}
\end{figure}

The vector potential from Eq.~(\ref{eqn: third expression of A}) has been calculated for a small rigid chiral molecule, (S)-d$_2$-oxirane shown in Fig.~\ref{fig:S-d2-oxirane}.
As will be clear from the numerical results, the vector potential contributes only a small fraction to the rotational strengths $R_k$ (with $k=1,\ldots,15$ for the (S)-d$_2$-oxirane and oxirane), as it is computed within a perturbation theory approach. The vector potential is first-order in the perturbation parameter $\boldsymbol{\lambda}_\nu(\mathbf R,t)$ and it appears as an explicit $\mathcal O(M_\nu^{-1})$ term in the expressions of the current and of the magnetic dipole moment. Further analysis, currently under investigation, is focussing on the calculation of corrections due to the vector potential in explicit non-adiabatic molecular dynamics, in order to estimate the actual effect of the vector potential on observable properties as the VCD signal.

\begin{figure}[ht!]
 \centering
 \includegraphics*[width=0.3\linewidth,angle=270]{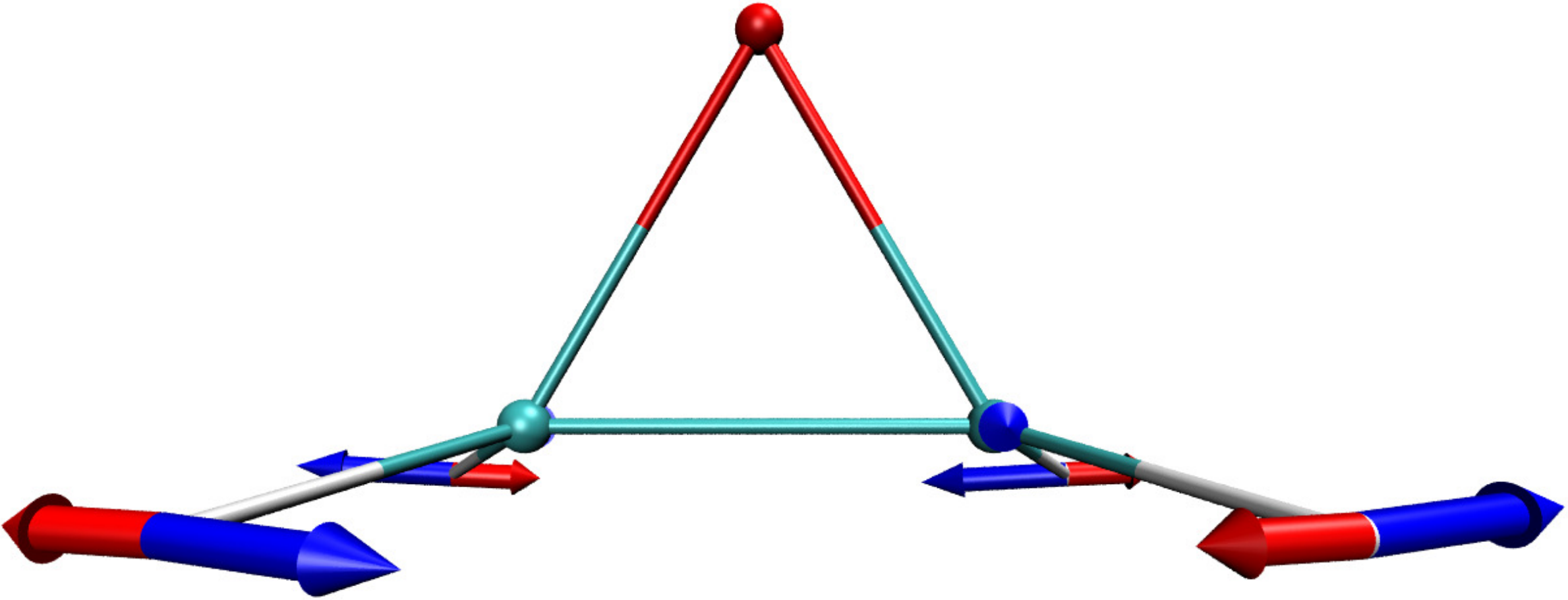}
 \includegraphics*[width=0.3\linewidth,angle=270]{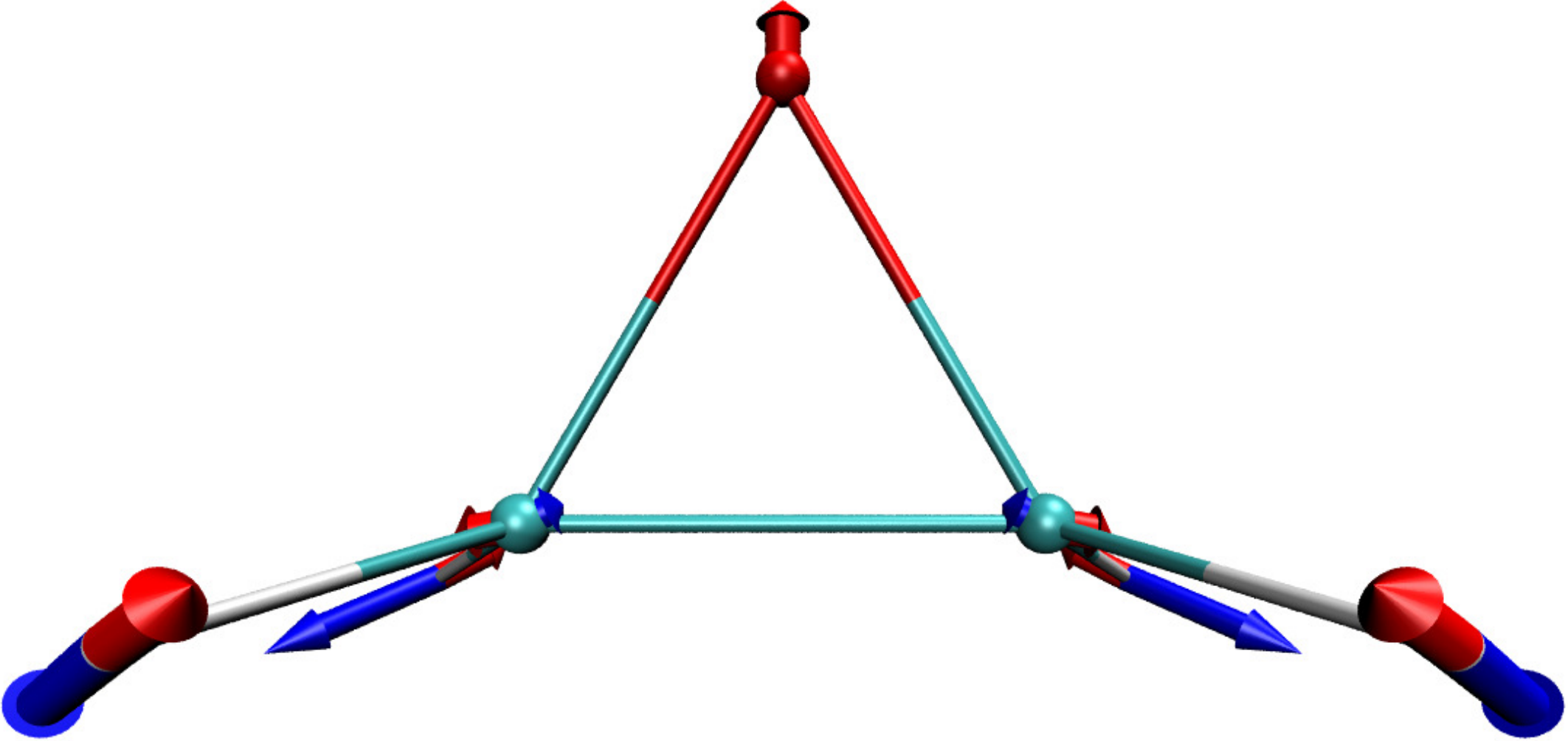}
 \caption{Vibrational modes at 1127 $\text{cm}^{-1}$ (upper panel) and at 1489 $\text{cm}^{-1}$ (lower panel) for oxirane, with nuclear velocities indicated as blue arrows. The corresponding vector potential is shown as red arrows.}
 \label{fig:oxirane-modes}
\end{figure}

Before presenting the results for (S)-d$_2$-oxirane, let us first discuss the case of oxirane, a non-chiral molecule. Oxirane differs from (S)-d$_2$-oxirane in the deuterium atoms, that are replaced by hydrogen atoms. In Fig.~\ref{fig:oxirane-modes} we draw as blue arrows~\footnote{The blue arrows are the velocities of the corresponding modes, i.e. removing the mass weighting of the eigenvectors.} the velocities corresponding to normal modes at 1127 $\text{cm}^{-1}$ (upper panel) and at 1489 $\text{cm}^{-1}$ (lower panel), which have been selected as examples among the 15 total modes. Perturbations parallel to these velocities are used in Eq.~(\ref{eqn: third expression of A}) to construct the vector potential, which is shown as red arrows in the figure.
It is very interesting to notice that in the case of a non-chiral system the vector potential maintains the same symmetry of the vibrational modes and is nearly anti-parallel to the nuclear displacement: this is what one would expect, if the vector potential is not to affect the VCD properties, i.e. current and magnetic dipole, and thus the rotational strength, of the molecule.

\begin{figure}[ht!]
 \centering
 \includegraphics*[width=0.3\linewidth,angle=270]{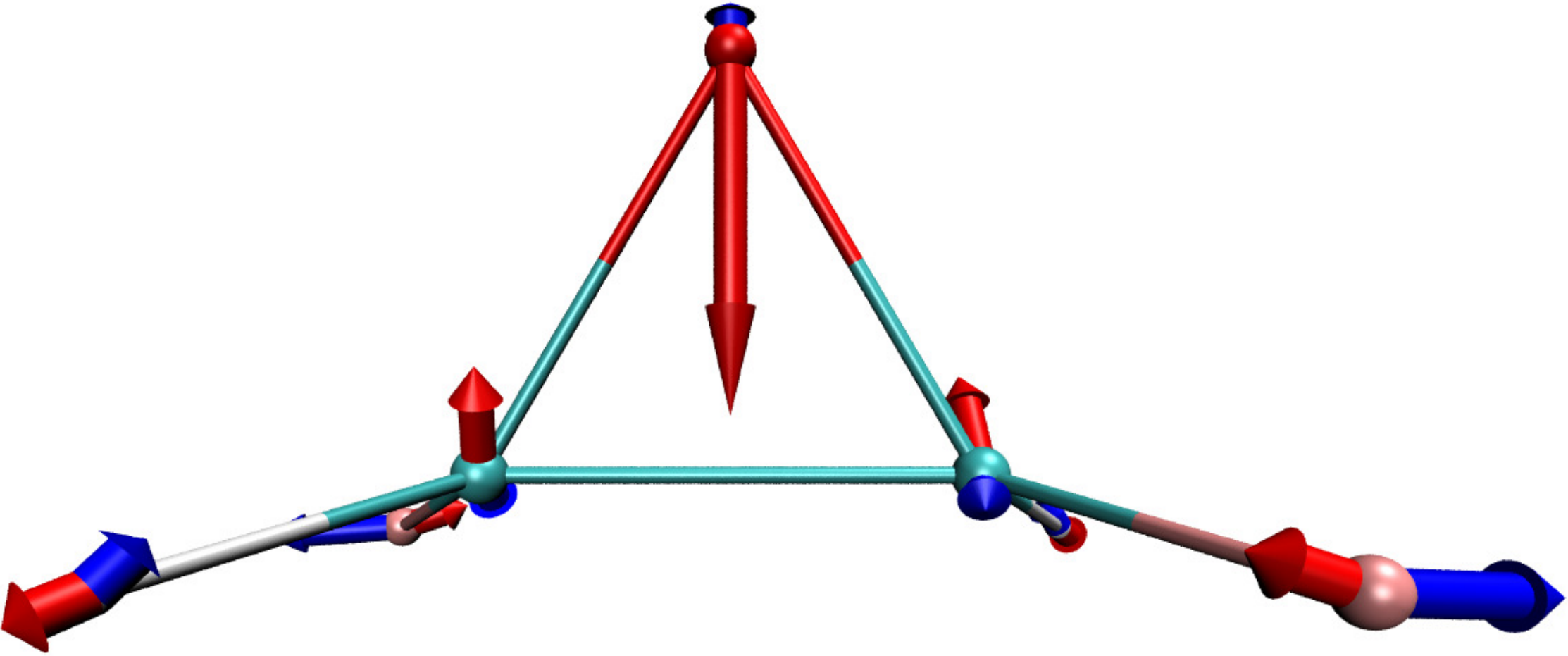}
 \includegraphics*[width=0.3\linewidth,angle=270]{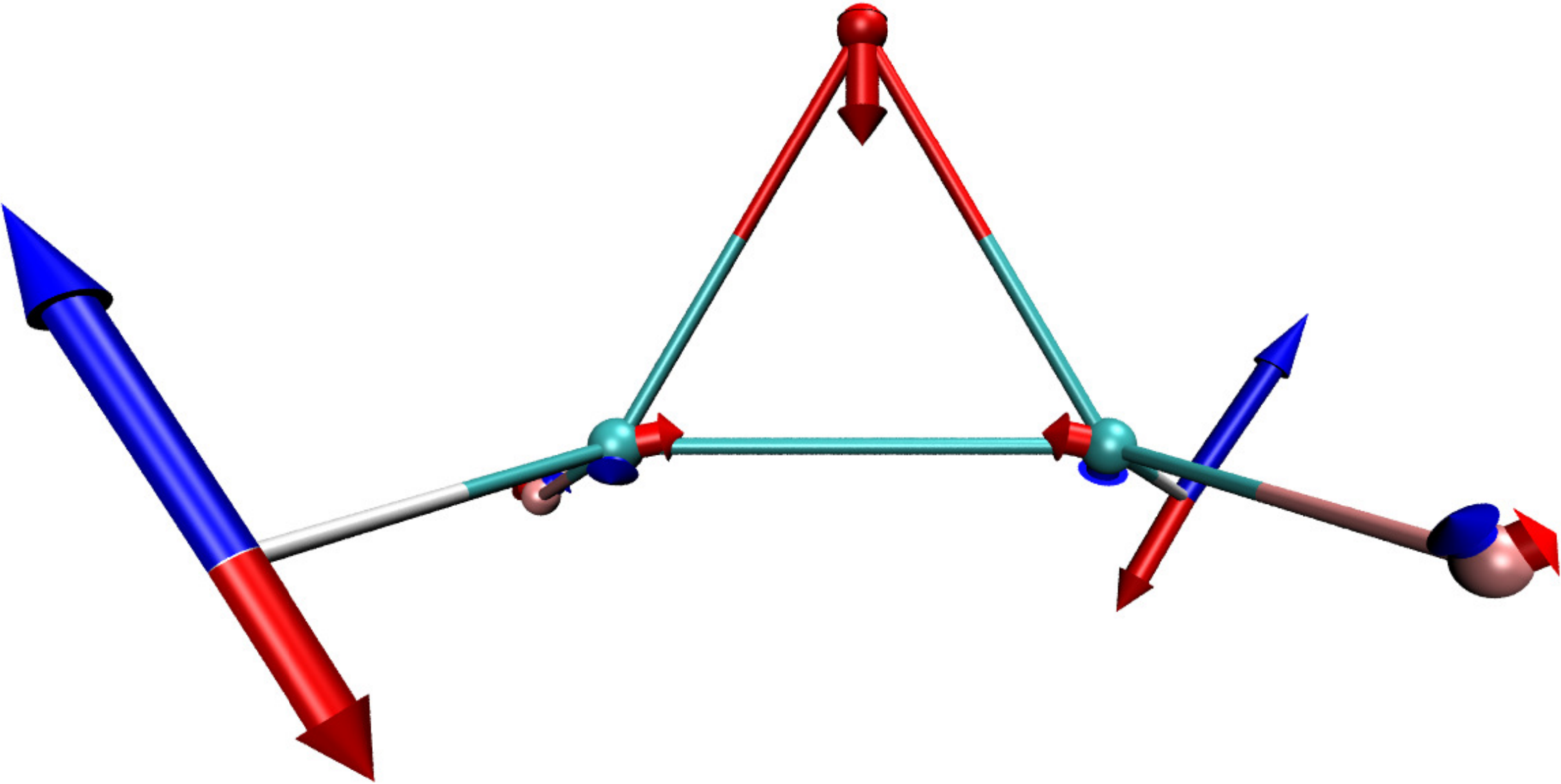}
 \caption{Vibrational modes at 896 $\text{cm}^{-1}$ (upper panel) and at 1089 $\text{cm}^{-1}$ (lower panel) for (S)-d$_2$-oxirane, with nuclear velocities indicated as blue arrows. The corresponding vector potential is shown as red arrows.}
 \label{fig:S-d2-oxirane-modes}
\end{figure}

In the case of (S)-d$_2$-oxirane, the results are quite different, as shown in Fig.~\ref{fig:S-d2-oxirane-modes}. Again, the velocities corresponding to the normal modes are indicated as blue arrows, whereas the vector potential is drawn in red. The selected modes are at 896 $\text{cm}^{-1}$ and at 1089 $\text{cm}^{-1}$. It is clear in this case that (i) a well-defined symmetry of the vector potential cannot be identified and, as a consequence, (ii) it is not simply (anti-)parallel to the normal modes velocities, as was the case for oxirane. This behavior thus results in an actual contribution of the vector potential to the VCD properties of (S)-d$_2$-oxirane.
Such contribution is quantitatively estimated by calculating the correction the the rotational strengths in Eq.~(\ref{eqn: dip_str_pos 2}) of (S)-d$_2$-oxirane, due to the vector potential terms in Eqs.~(\ref{eqn: electric dipole}) and~(\ref{eqn: magnetic dipole}). Table~\ref{tab: S-d2-oxirane} lists, for all modes in the (S)-d$_2$-oxirane, these rotational strengths $R_k$ and the corrections $\Delta R_k$ due to the presence of the vector potential in the current and in the magnetic dipole moment.

\begin{table}
 \begin{center}
 \caption{Normal modes for (S)-d$_2$-oxirane. The frequencies of the modes are indicated in the first column, the rotational strengths $R$ are listed in the second column, from Eq.~(\ref{eqn: dip_str_pos 2}), the corrections $\Delta R$ due to the vector potential are reported in the third (absolute value) and fourth (relative correction) columns.}
 \begin{tabular}{rrrr}
  \hline
  \hline
  \multicolumn{1}{c}{$\tilde{\nu}$} &
  \multicolumn{1}{c}{$R$} &
  \multicolumn{1}{c}{$\Delta R$} &
  \multicolumn{1}{c}{$\Delta R/R$} \\
  \multicolumn{1}{c}{($\text{cm}^{-1}$)} &
  \multicolumn{1}{c}{($10^{-44}\text{esu}^2 \text{cm}^2$)} &
  \multicolumn{1}{c}{($10^{-44}\text{esu}^2 \text{cm}^2$)} &
  \multicolumn{1}{c}{(\%)} \\
  \hline
  \hline
   647.50  &  -0.45  & -0.003 & 0.67 \\
   733.42  &  10.54  &  0.016 & 0.15 \\
   769.76  &   3.29  &  0.001 & 0.05 \\
   856.38  &   2.70  &  0.002 & 0.09 \\
   894.67  &  -3.89  &  0.006 & 0.15 \\
   936.33  & -20.26  &  0.001 & 0.01 \\
  1088.21  &   8.34  & -0.027 & 0.32 \\
  1093.95  &  -4.97  &  0.004 & 0.09 \\
  1210.44  &   10.45  & -0.029 & 0.28 \\
  1326.86  &  -0.76  & 0.0002 & 0.03 \\
  1377.38  &  -8.17  &  0.025 & 0.31 \\
  2235.16  & -22.90  & -0.010 & 0.04 \\
  2244.19  &  16.78  &  0.011 & 0.07 \\
  3047.68  & -32.59  & -0.063 & 0.19 \\
  3054.15  &  47.04  &  0.047 & 0.10 \\
  \hline
  \hline
 \end{tabular}\label{tab: S-d2-oxirane}
 \end{center}
\end{table}

As discussed above, we notice from the results reported in Table~\ref{tab: S-d2-oxirane} that, despite the fact that the vector potential is non-zero, its effect is quite small, being of the order $\mathcal O(M^{-1}_\nu)$. In fact, while the $M^{-1}_\nu$ dependence in Eqs.~(\ref{eqn: electric dipole}) and~(\ref{eqn: magnetic dipole}) is removed in the first contributions, being these first terms proportional to the momentum, the second terms are actually $\mathcal O(M^{-1}_\nu)$. We recall, however, that in the procedure developed in this paper, the vector potential is evaluated within the NVPT, thus being first-order in the perturbation. In a situation where the electronic wave function has a strong non-adiabatic character, namely where the correction to a BO-type wave function is not small in the nuclear velocity, a larger contribution may be expected. Moreover, in the cases where the vector potential is singular, e.g. for adiabatic states that are locally degenerate in $\mathbf R$-space, this correction may become very important. However, further studies are required to develop a scheme that allows for the calculation of the vector potential beyond the NVPT.

\subsection{Comparison with magnetic field perturbation theory}\label{sec: MFPT vs. NVPT}
\begin{figure}[ht!]
 \centering
 \includegraphics*[width=0.3\linewidth]{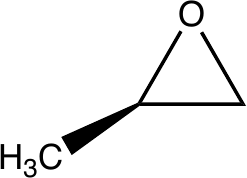}
 \caption{(R)-propylene-oxide.}
 \label{fig:R-propylene-oxide}
\end{figure}

\begin{figure}[ht!]
 \centering
 \includegraphics*[width=0.3\linewidth]{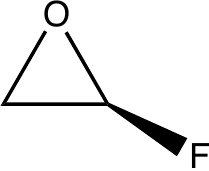}
 \caption{(R)-fluoro-oxirane.}
 \label{fig:R-fluoro-oxirane}
\end{figure}

Further molecular systems have been investigated, namely (R)-propylene-oxide and (R)-fluoro-oxirane shown in Figs.~\ref{fig:R-propylene-oxide} and~\ref{fig:R-fluoro-oxirane}.

In this section we report the dipole $D$ and rotational $R$ strengths calculated by employing NVPT, indicated with the symbols $D_{\text{NVP}}$ and $R_{\text{NVP}}$ in Tables~\ref{tab: S-d2-oxirane bis},~\ref{tab: RPO} and~\ref{tab: RFO}, and we compare these results with the magnetic field perturbation (MFP) theory~\cite{Stephens_JPC1985}, $D_{\text{MFP}}$ and $R_{\text{MFP}}$ in the tables. Such comparison has been carried out also for (S)-d$_2$-oxirane (Table~\ref{tab: S-d2-oxirane bis}). Furthermore, Tables~\ref{tab: RPO} and~\ref{tab: RFO} show the corrections $\Delta R$ to the rotational strengths due to the vector potential term in Eq.~(\ref{eqn: third expression of A}), as already presented for the case of (S)-d$_2$-oxirane in Section~\ref{sec: examples}. In all tables the first column indicates the normal modes frequency, the second and third columns are the dipole strengths from MFP and NVP theories, the forth and fifth columns show the rotational strengths from MFP and NVP theories. In Tables~\ref{tab: RPO} and~\ref{tab: RFO} the sixth and seventh columns are the corrections computed from Eq.~(\ref{eqn: third expression of A}), which in general are the same order of magnitude as the corrections reported in Table~\ref{tab: S-d2-oxirane} for (S)-d$_2$-oxirane.

\begin{table}
 \begin{center}
 \caption{Normal modes, dipole and rotational strengths, for (S)-d$_2$-oxirane.}
 \begin{tabular}{r|rr|rr}
  \hline
  \hline
  \multicolumn{1}{c|}{$\tilde{\nu}$} &
  \multicolumn{1}{c}{$D_\text{MFP}$} &
  \multicolumn{1}{c|}{$D_\text{NVP}$} &
  \multicolumn{1}{c}{$R_\text{MFP}$} &
  \multicolumn{1}{c}{$R_\text{NVP}$} \\
  \multicolumn{1}{c|}{($\text{cm}^{-1}$)} &
  \multicolumn{2}{c|}{($10^{-40}\text{esu}^2 \text{cm}^2$)} &
  \multicolumn{2}{c}{($10^{-44}\text{esu}^2 \text{cm}^2$)} \\
  \hline
  \hline
   647.50 &   0.55 &   0.85 &  -0.35 &  -0.45  \\
   733.42 & 123.35 & 124.88 &   8.73 &  10.54  \\
   769.76 &  53.44 &  51.77 &   3.17 &   3.29 \\
   856.38 & 145.31 & 145.55 &   4.31 &   2.70 \\
   894.67 &   9.78 &  10.24 &  -3.37 &  -3.89  \\
   936.33 &  39.73 &  39.24 & -19.14 & -20.26  \\
  1088.21 &   3.79 &   4.44 &   6.95 &   8.34  \\
  1093.95 &   1.41 &   1.71 &  -3.98 &  -4.97 \\
  1210.44 &  26.26 &  26.09 &   9.56 &  10.45 \\
  1326.86 &   0.34 &   0.37 &  -0.91 &  -0.76  \\
  1377.38 &  11.65 &  10.78 &  -7.50 &  -8.17 \\
  2235.16 &  49.17 &  50.88 & -22.60 & -22.90  \\
  2244.19 &  12.63 &  12.81 &  16.80 &  16.78 \\
  3047.68 &  11.43 &  11.66 & -32.80 & -32.59  \\
  3054.15 &  58.64 &  60.16 &  46.63 &  47.04  \\
  \hline
  \hline
 \end{tabular}\label{tab: S-d2-oxirane bis}
 \end{center}
\end{table}
\begin{table}
 \begin{center}
 \caption{Normal modes, dipole and rotational strengths (with corrections), for (R)-propylene-oxide. $\Delta R$ and $\Delta R/R$ in the last columns are indicated in $10^{-44}$esu$^2$cm$^2$ and \%, respectively.}
 \begin{tabular}{r|rr|rr|rr}
  \hline
  \hline
  \multicolumn{1}{c|}{$\tilde{\nu}$} &
  \multicolumn{1}{c}{$D_\text{MFP}$} &
  \multicolumn{1}{c|}{$D_\text{NVP}$} &
  \multicolumn{1}{c}{$R_\text{MFP}$} &
  \multicolumn{1}{c|}{$R_\text{NVP}$} &
  \multicolumn{1}{c}{$\Delta R$} &
  \multicolumn{1}{c}{$\Delta R/R$} \\
  \multicolumn{1}{c|}{$(\text{cm}^{-1})$} &
  \multicolumn{2}{c|}{($10^{-40}\text{esu}^2 \text{cm}^2$)} &
  \multicolumn{2}{c|}{($10^{-44}\text{esu}^2 \text{cm}^2$)} &
  \multicolumn{1}{c}{} &
  \multicolumn{1}{c}{} \\
  \hline
  \hline
 202.12 &    6.91 &   7.12 &    3.54 &   3.47 & -0.001 & 0.03 \\
 355.33 &   45.22 &  46.63 &  -12.84 & -12.56 & 0.008 & 0.06 \\
 398.40 &   38.74 &  38.86 &   -3.72 &  -3.79 & 0.004 & 0.11 \\
 717.36 &   55.11 &  52.51 &   13.88 &  13.21 & 0.005 & 0.04 \\
 795.26 &  217.23 & 219.15 &    2.47 &   1.62 & 0.007 & 0.44 \\
 875.72 &   18.54 &  17.25 &   26.36 &  26.99 & 0.039 & 0.14 \\
 929.21 &   51.55 &  51.53 &  -35.52 & -37.03 & -0.049 & 0.13 \\
1008.29 &   24.84 &  26.62 &    2.88 &   4.53 & -0.004 & 0.09 \\
1089.09 &   18.53 &  19.17 &   -6.03 &  -6.56 & 0.006 & 0.09 \\
1112.88 &    7.92 &   7.79 &    6.65 &   7.50 & 0.023 & 0.31 \\
1126.68 &   11.68 &  12.56 &  -13.44 & -14.67 & -0.034 & 0.23 \\
1150.27 &    1.51 &   1.40 &    1.54 &   1.23 & 0.003 & 0.24 \\
1246.96 &   19.77 &  19.85 &   -8.06 &  -8.01 & -0.004 & 0.05 \\
1371.08 &   10.35 &   9.83 &    3.30 &   3.53 & 0.007 & 0.19 \\
1388.57 &   60.08 &  60.10 &   13.99 &  15.15 & 0.007 & 0.05 \\
1447.69 &   13.15 &  14.16 &    1.34 &   1.45 & 0.005 & 0.32 \\
1461.62 &   15.41 &  16.62 &   -1.69 &  -1.90 & -0.008 & 0.42 \\
1480.79 &   10.14 &   9.99 &    4.66 &   4.69 & -0.005 & 0.11 \\
2955.51 &   27.68 &  28.86 &    1.64 &   1.64 & 0.0002 & 0.01 \\
3000.54 &   41.29 &  44.50 &   -0.29 &   0.20 & -0.009 & 4.53 \\
3005.59 &   22.70 &  24.14 &    5.13 &   6.04 & -0.034 & 0.56 \\
3007.28 &   22.57 &  22.86 &  -13.92 & -15.14 & 0.053 & 0.35 \\
3032.47 &   47.06 &  50.07 &    7.29 &   7.16 & -0.019 & 0.27 \\
3079.38 &   41.31 &  41.09 &   -7.19 &  -7.31 & 0.013 & 0.17 \\
  \hline
  \hline
 \end{tabular}\label{tab: RPO}
 \end{center}
\end{table}
\begin{table}
 \begin{center}
 \caption{Normal modes, dipole and rotational strengths (with corrections), for (R)-fluoro-oxirane.  $\Delta R$ and $\Delta R/R$ in the last columns are indicated in $10^{-44}$esu$^2$cm$^2$ and \%, respectively.}
 \begin{tabular}{r|rr|rr|rr}
  \hline
  \hline
  \multicolumn{1}{c|}{$\tilde{\nu}$} &
  \multicolumn{1}{c}{$D_\text{MFP}$} &
  \multicolumn{1}{c|}{$D_\text{NVP}$} &
  \multicolumn{1}{c}{$R_\text{MFP}$} &
  \multicolumn{1}{c|}{$R_\text{NVP}$} &
  \multicolumn{1}{c}{$\Delta R$} &
  \multicolumn{1}{c}{$\Delta R/R$} \\
  \multicolumn{1}{c|}{$(\text{cm}^{-1})$} &
  \multicolumn{2}{c|}{($10^{-40}\text{esu}^2 \text{cm}^2$)} &
  \multicolumn{2}{c|}{($10^{-44}\text{esu}^2 \text{cm}^2$)} &
  \multicolumn{1}{c}{} &
  \multicolumn{1}{c}{} \\
  \hline
  \hline
  411.61 &   52.63 &    53.11 &   9.48 &   9.80 & -0.003 & 0.03 \\
  482.91 &   30.46 &    31.23 &  -3.10 &  -2.91 & 0.002 & 0.06 \\
  733.56 &  124.68 &   123.64 &  40.79 &  39.91 & 0.031 & 0.08 \\
  804.61 &  501.12 &   497.82 & -12.79 &  -9.85 & 0.007 & 0.07 \\
  927.57 &  244.98 &   246.52 & -27.57 & -34.46 & -0.052 & 0.15 \\
 1059.05 &   28.75 &    25.77 &  -9.38 &  -9.09 & -0.019 & 0.21 \\
 1069.68 &  312.66 &   321.09 &  22.55 &  22.47 & 0.048 & 0.21 \\
 1106.47 &    5.52 &     4.95 &  -8.37 &  -8.84 & -0.022 & 0.25 \\
 1125.52 &   11.52 &    11.28 &   4.11 &   4.77 & 0.005 & 0.10 \\
 1252.78 &   88.68 &    87.54 &  -0.07 &   1.73 & 0.006 & 0.32 \\
 1344.65 &  150.66 &   150.18 &  -6.39 &  -7.04 & -0.016 & 0.23 \\
 1470.12 &   42.55 &    44.05 &   0.73 &   0.77 & 0.008 & 1.06 \\
 3024.87 &   20.22 &    21.53 &   1.64 &   1.53 & 0.003 & 0.16 \\
 3068.24 &   22.58 &    23.21 &  -1.07 &  -1.00 & 0.008 & 0.84 \\
 3115.60 &   14.50 &    14.02 &   0.34 &   0.37 & -0.007 & 1.79 \\
  \hline
  \hline
 \end{tabular}\label{tab: RFO}
 \end{center}
\end{table}

From the comparison between the two perturbation approaches, we notice an overall very good agreement not only in the absolute values of the dipole and rotational strengths, but also in the signs of the rotational strengths for the three systems investigated here. The MFP theory of Stephens~\cite{Stephens_JPC1985} can be considered a ``more standard'' approach, nowadays implemented in most quantum-chemistry packages, thus it represents a suitable benchmark for the new approach introduced in Ref.~\cite{Scherrer_JCTC2013} and discussed in the present work.

\section{Conclusions}\label{sec: conclusions}
One of the main goal of the paper has been to provide rigorous basis for the development of the NVPT approach to VCD. In this context, the complete adiabatic approach proposed by Nafie~\cite{Nafie_JCP1983} was adopted in previous study~\cite{Scherrer_JCTC2013} as starting point, where the electron-nuclear wave function is approximated as a single product of a (nuclear) vibrational contribution and an electronic term. In particular, such electronic term contains corrections to the BO state which are first-order in the nuclear velocity. In the present work, we make this idea \textit{exact}, in the sense that the starting point is not an approximate factorized form of the full wave function. The starting point is provided by the exact factorization of the electron-nuclear wave function, where approximations are inserted at a later stage in order to make numerical calculations feasible. The method outlined here can thus be seen as a rigorous basis for NVPT: at the first stage of the derivation we describe how to recover the BO working equation from the exact electronic equation and at the second stage a perturbation to BO is considered. Also, this perturbation does not rely on the use of the nuclear velocity as small parameter, in fact such parameter is, more generally, related to the spatial variations of the nuclear wave function from the factorization. Only in the classical limit, at $\mathcal O(\hbar^0)$, these variations lead to an interpretation in terms of nuclear velocity. In the new approach presented here, a full quantum picture can be maintained, without invoking the classical approximation.

The second main result confirms the importance of using the exact factorization as starting point for the development of approximations. The time-dependent vector potential of the theory naturally appears in the observables, i.e. the current and the magnetic dipole moment, necessary for the calculation of the VCD spectrum. Therefore, within the perturbation approach presented in the paper, we have evaluated the vector potential using the harmonic approximation for the nuclear motion. In this case, the contribution has been shown to be small, but only further investigation, for instance in the context of non-adiabatic molecular dynamics, will clarify the actual extent of non-adiabatic corrections to the VCD signal. Also, situations where the non-adiabatic couplings are important shall be investigated, for instance for low-lying excited states~\cite{Nafie_JPCA2004}, where the exact factorization approach offers a strategy to overcome the limitations of BO approximation in a rigorous way.

According to the procedure presented in this work, NVPT is suitable for an implementation in any ab-initio molecular dynamics code. Therefore, NVPT can be easily employed for the study of VCD properties of chiral molecules in solutions and for direct comparison with experimental data. Such procedure allows also to evaluate the corrections due the vector potential from the exact factorization approach. As it requires a DFPT calculation for each geometry sampled by the molecular dynamics trajectory, the numerical cost of a NVPT calculation is slightly larger than standard BO molecular dynamics but indeed feasible.

\section*{Acknowledgements}
The authors would like to thank Axel Schild for his help in improving the presentation of the paper. Partial support from the Deutsche Forschungsgemeinschaft (SFB 762) and from the European Commission (FP7-NMP-CRONOS) is gratefully acknowledged.

\appendix
\section{Invariance under choice of the origin}\label{app: origin dependence}
One of the main problems connected to the evaluation of molecular properties and spectroscopies depending on the magnetic field is to assure origin invariance of the final results. In case of VCD, this requires the evaluation of the electric and magnetic dipole moments, or accordingly the APT and the AAT. While the APT shows no origin dependency, the exact AAT transforms under shifts of the origin $\boldsymbol{\mathcal{O}} = \boldsymbol{\mathcal{O}}' + \boldsymbol{\Delta}$ as
\begin{align}
 \mathcal{M}^{\nu \mathcal{O}}_{\alpha \beta} &= \mathcal{M}^{\nu \mathcal{O}'}_{\alpha \beta} 
 - \sum_{\gamma \delta} \tfrac{1}{2c} \epsilon_{\beta \gamma \delta} \Delta_\gamma \mathcal{P}^{\nu}_{\alpha\delta} \label{eq:exact_aat_translation_relation}.
\end{align}
The rotational strength as a physical observable is gauge invariant
\begin{align}
 R_k &= \sum_{\alpha \alpha' \beta} \sum_{\nu \nu'} \mathcal{P}^{\nu}_{\alpha\beta} \mathcal{M}^{\nu'\mathcal{O}'}_{\alpha'\beta} S^\nu_{\alpha k} S^{\nu'}_{\alpha' k} \nonumber \\
           & - \sum_{\alpha \alpha' \beta \gamma \delta} \sum_{\nu \nu'} \tfrac{1}{2c} \epsilon_{\beta \gamma \delta} \Delta_\gamma \mathcal{P}^{\nu}_{\alpha\beta} \mathcal{P}^{\nu'}_{\alpha'\delta} S^\nu_{\alpha k} S^{\nu'}_{\alpha' k},
\end{align}
since the second terms constitute triple products containing two identical vectors.

The evaluation of origin dependent operators under periodic boundary conditions has been extensively discussed in the literature~\cite{Resta1994, Resta2010, Nafie-2011}. A convenient approach is the combination of statewise origins with maximally localized Wannier orbtials, which has been applied successfully to the calculation of nuclear magnetic resonance chemical shifts~\cite{Berghold-2000, Sebastiani-2001}. The canonical $\phi_{o}$ and localized $\varphi_{o}$ states are mutually related via the unitary transformation for the unperturbed ground-state orbitals
\begin{align}
 |\varphi_o\rangle = \sum_{o'} U_{oo'}^{(0)} |\phi_{o'}\rangle.
\end{align}
This approach is based on the natural assumption that the response orbitals are sufficiently localized in the region of their respective unperturbed ground-state orbitals.
In the distributed origin (DO) gauge, the position operators are calculated with the corresponding Wannier center as its statewise origin
\begin{align}
 {\bf r}_o = \langle \varphi_o | \hat{\bf r}  |\varphi_o\rangle.
\end{align}
The electronic AAT in a statewise DO gauge then is given by
\begin{align}
 \big(\mathcal{I}^\nu_{\alpha \beta} \big)_{\scriptscriptstyle \text{DO}}^o = \frac{e}{mc} \langle \varphi_o | (\hat{r}_\gamma - r_{o\gamma}) \hat{p}_\delta \epsilon_{\beta\gamma\delta} | \varphi^{(1)}_{o,(\nu,\alpha)} \rangle
\end{align}
and can be translated back to the common origin form via
\begin{align}
 \mathcal{I}^{\nu \mathcal{O}}_{\alpha \beta} &= \sum_{o} \big(\mathcal{I}^\nu_{\alpha \beta} \big)_{\scriptscriptstyle \text{DO}}^o \nonumber \\
 &+ \sum_{o\gamma \delta}  \tfrac{1}{2c} \epsilon_{\beta \gamma \delta} (r_{o\gamma}- \mathcal{O}_\gamma) \mathcal{E}^{\nu o}_{\alpha\delta} \label{eq:aat_statewise_dog}.
\end{align}
where $\mathcal{E}^{\nu o}_{\alpha\delta}$ is the contribution of the state $o$ to the electronic APT. The numerical results in a supercell calculation are the same for canonical and Wannier orbitals~\cite{Scherrer_JCTC2013}.


%

\end{document}